\documentclass[doublecol]{epl2} 

\usepackage{amsmath}
\usepackage{isotope}
\usepackage{float}
\usepackage{placeins}

\title{Parasitic neutron beam monitoring: proof of concept on gamma monitoring of neutron chopper phases}
\shorttitle{Parasitic neutron beam monitoring} 
\author{F. Issa\inst{1} \and R. Hall-Wilton\inst{1,2} \and A. Quintanilla\inst{1}\and M. Olsson\inst{1}\and D. Zielinski\inst{1} \and K. Kanaki\inst{1} \and N. Tsapatsaris\inst{1}}
\shortauthor{F. Issa \etal}
\institute{
  \inst{1} European Spallation Source ESS ERIC, P.O Box 176, SE-221 00 Lund, Sweden\\
  \inst{2} Dipartimento di Fisica ``G. Occhialini'', Universit\`a degli Studi di Milano-Bicocca, Piazza della Scienza 3, 20126 Milano, Italy
}
\abstract{
Neutron beam monitors are an essential diagnostic component of neutron scattering facilities.
They are used to measure neutron flux, calibrating experiments performed on the instruments, allowing measurement of facility performance, understanding of the effect on the neutrons of beam-line components (such as choppers),  calibration of detectors and tracking of beam stability.
Ideally beam monitors {should} not perturb the beam.
Previous work shows commercial  beam monitors attenuate the beam by a few percent in the worst case due to the 1-2 mm thick Aluminium entrance and exit windows and the material inside.
Parasitic methods of neutron beam diagnostics, where there is no beam monitor directly in the beam, would be preferable.
This paper presents the concept of a parasitic method of monitoring the beam which can be used for neutron chopper phasing.
This is achieved by placing a gamma detector close to a rotating chopper and measures a signal proportional to the flux absorbed by the chopper.
Neutrons interact with the Boron absorber on the chopper disc lead to gamma emission at 480 keV.
Detection of these gamma rays is used to determine the chopper phasing and timing.
Potentially information on the flux of the beamline can be extracted.
Results from a proof of concept implementation show that diagnosis of neutron chopper phases is feasible.
}
\begin{document}
\maketitle
\section{Introduction }

  Neutrons from  spallation sources such as at SNS ~\cite{SNS}, ISIS ~\cite{ISIS} and J-PARC ~\cite{JPARC} and ESS˜\cite{ESS} are created by a primary proton beam accelerated with a specific repetition rate toward a heavy target which can be tungsten, lead or mercury.
  The produced neutrons are created with high energy (up to ca. GeV); the neutron's energy is  reduced by a moderator made of light elements such as hydrogen.
  The neutrons are then transported from the moderator to the sample position by neutron guides.
  As the neutron's energy defines its velocity (about 2200m/s at thermal energies of 25meV which is equivalent to a wavelength of about 1.8 Angstrom), the time-of-flight of the neutron can be used to select the neutron energy.
  This selection could be achieved by several sets of mechanical {\it ``choppers"}˜\cite{choppers}. Choppers are mechanical rotating discs, with slits which define the timing of the neutrons which are allowed to proceed down the beamline. Typically several chopper pairs are used per instrument proposed for the European Spallation Source (ESS) facility˜\cite{ESS-instrumentsuite}. The chopper system and the length of the instrument control the neutron wavelength and flux. The chopper discs are coated with a neutron absorbent material such as Gd or B in order to stop the unwanted neutrons.

  The length of a neutron instrument varies from  {a} few metres up to 160 metres.
  A set of beam monitors is needed per instrument in order to diagnose its main components such as choppers and guides section and to determine flux on sample.

  The requirements for the monitors vary greatly with respect to their location and purpose. Several types of monitors are needed to fulfil all these requirements.
  In recently built instruments, and in particular for instruments planned for the ESS facility, the number of beam line components and their complexity is increasing.
  This is especially the case for neutron choppers, where the number of choppers planned in the baseline instrument suite  {for ESS (approximately 150) is} comparable to the number of choppers currently installed  {worldwide}˜ {\cite{Maulerova2020}}.
  This in turn implies an increased need in beam-line diagnostics for efficient commissioning, operation and understanding of the instrument.

\begin{figure}
\centering
\includegraphics[width=70mm]{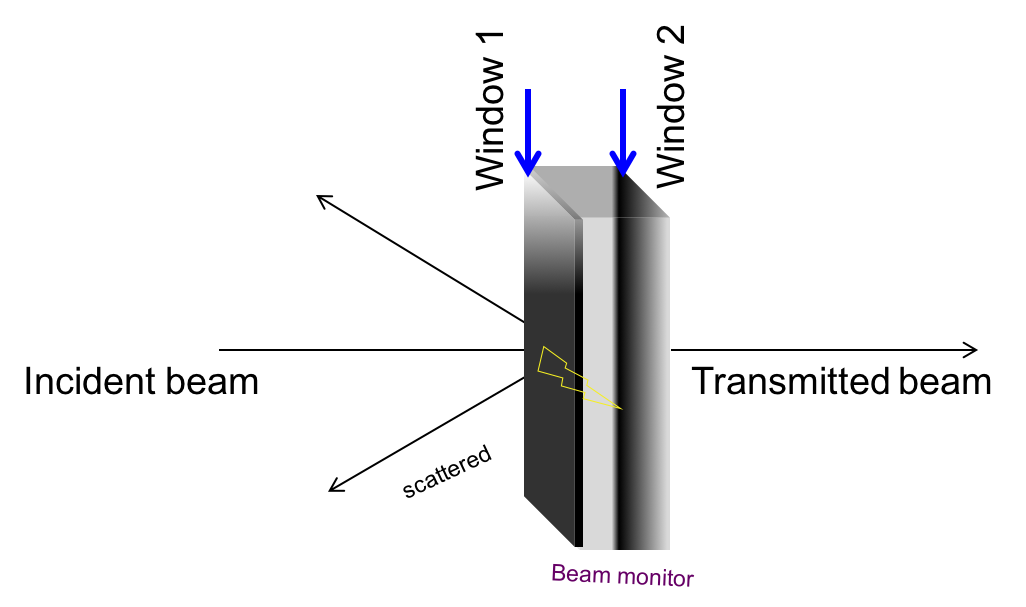}
\caption{Typical beam monitor in the direct beam showing the entrance and exit window.
\label{BM_att}}
\end{figure}

	Typically  neutron beam monitors are simple neutron detectors with sufficiently low efficiency (10$^{-6}$ -10$^{-1}$) so that a low percentage of the incoming  beam is absorbed or scattered as shown in fig.~\ref{BM_att}. They are used to ensure that the neutron flux, beam distribution,  and pulse timing correspond to those expected from the design of the instrument. In addition, they are used to determine the neutron flux at the sample in order to correctly interpret the scattering data.  Different types of beam monitors from a variety of suppliers have been characterised in previous work  showing a high attenuation factor for most of the monitors mainly due to the entrance and exit window~\cite{BMpaper} as shown in fig.~\ref{BM_att}.

	The  {desideratum} is a parasitic method of beam monitoring that avoids attenuation of the beam.
One such quasi-parasitic method has recently been investigated˜\cite{VanadiumBM}.
  The parasitic concept developed here takes advantage of a chopper disc coated with Boron carbide.  Neutrons blocked by this disc interact with  Boron leading to gamma emission at 480 keV based on the interaction shown below:
%

\begin{equation}
\label{eq1}
\isotope{n} + \isotope[10]{B} \to \isotope[7]{Li} + \alpha \ (6\%) \
\end{equation}

\begin{equation}
\label{eq2}
\isotope{n} + \isotope[10]{B}  \to \isotope[7]{Li} + \alpha+ \gamma(0.48\,\mathrm{MeV}) \ (94\%) \
\end{equation}

\begin{figure}
\centering
\includegraphics[width=60mm]{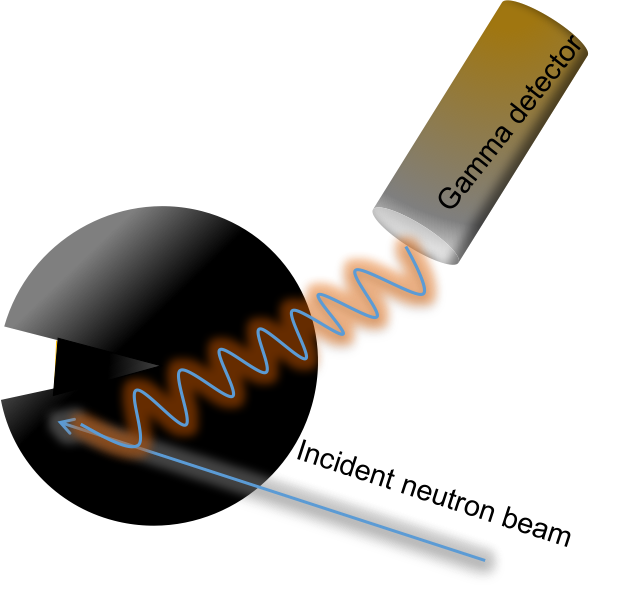}
\caption{Cartoon showing a chopper disc coated with Boron with a gamma detector placed away from the incident beam. The incident thermal neutron beam can interact with Boron on the disc and lead to gamma emission which is then detected by a gamma detector.
\label{Concept}}
\end{figure}

The emitted gamma rays can be measured using a detector which could be a scintillator made of e.g. NaI or  {LaBr$_{3}$}. This can be placed close to the chopper disc and therefore requires no additional material in the direct neutron beam;  thus no perturbation of the beam occurs.
The concept is shown in fig.~\ref{Concept}.

 {One of the challenges of this measurement is that Boron is a very commonly used material on neutron beamlines: it is used for shielding, for slits and collimators and is often present in the substrate material of neutron guides.
Therefore the 480 keV gamma rays are ubiquitious around neutron instruments; the challange of this concept is to show that the signal from the chopper can be seen above the background from the rest of the facility.}

\section{Concept of Measurement}

For a traditional neutron beam monitor,  {the transmitted flux, $I$,  is:

\begin{equation}
\label{eq3}
I(\lambda) = I_0(\lambda) - S(\lambda) - A(\lambda) - M (\lambda)
\end{equation}

where $I_0$ is the neutron intensity incident} on the beam monitor.
$S$ and $A$ are the neutrons scattered from the beam monitor and those absorbed in the non-sensitive material of the beam monitor respectively.
 {$M$ is the measured signal in the beam monitor.}
These neutrons are not transmitted and are lost from the beam.
It should be noted that, experimentally, the difference between scattered and measured neutrons is one of detector geometry, as has been discussed elsewhere\cite{Klausz2019}.
$\lambda$ is the neutron wavelength.

The measured flux is thus dependent upon the neutron wavelength.
There is an additional correction from the fraction absorbed in dead material and scattered out of the beamline, which can be sizable˜\cite{BMpaper}.
The corrections for wavelength are calculable and good design can reduce the attenuation of the incident beam to the percent level.
However these corrections represent complications to the measurement schema; often these are neglected.

In contrast, the concept of measuring the gamma rays emitted relies on measuring neutrons absorbed in the Boron of the neutron chopper disc.
This is actually the incident flux which is {\it not} transmitted by the chopper, or  {$\overline I$.}
The neutron chopper is a rotating mechanical device and therefore has a time varying function of neutron transmission.
This means that its transmission function can be expressed as:

 {
\begin{equation}
\label{eq5}
I_0(t) = A(t) + I(t)
\end{equation}
}

where  {$I_0$} is the neutron intensity incident on the chopper, $A$ is the neutrons absorbed in the chopper and  {$I$} is the neutrons transmitted down the neutron beamline.
Scattering from the  {chopper} components is neglected here.

The gamma detector measures the gammas emitted with an efficiency which can be expressed as:

\begin{equation}
\label{eq6}
M(t) = A(t) \cdot \epsilon_n(\lambda) \cdot
K_{BR}
\cdot \Omega \cdot \epsilon_\gamma (E_\gamma=480 keV)
\end{equation}

where $M$ is the detected  {signal} and $A$ is the neutrons absorbed in the chopper.
$\epsilon_n$ is the fraction of neutrons incident on the Boron Carbide coating of the chopper which are absorbed.
K$_{BR}$ is the branching ratio (94\%) for the gamma emmission fraction, given in equation\ ˜\ref{eq2}.
$\Omega$ is the fraction of solid angle subtended by the sensitive area of the gamma detector to the area where the neutron beam impinges on the chopper disc.
Lastly, $\epsilon_\gamma$ is the efficiency of the gamma detector to 480 keV gamma rays.

As neutron choppers aspire to a high
{\it ``blackness"} to neutrons when closed, the $\epsilon_n$ factor will, by design, be close to 1 and almost constant across the neutron wavelength of interest.
The other factors are not variable with neutron wavelength.
This means that the measurement efficiency is not variable with wavelength.

In summary, this concept should give information on absorbed flux on the beamline as a function of time.
In particular, it should give prompt time information on changes in flux, i.e. the opening and closing edges of chopper slits.

\section{Experimental Setup }
 A   {LaBr$_{3}$} gamma  {scintillator} from Saint Gobain~\cite{saintgobain} with  {a 1.5 x 1.5} inch scintillator crystal was used to detect the emitted gamma from a Boron interaction.
  {The scintillator was hermetically integrated (Saint Gobain Crystals 38S38/ 2) with a photomultiplier tube (PMT product number R6231-100-01 PMT) then the signal was shaped and amplified using the 885 dual spectroscopy amplifier from ORTEC~\cite{ORTEC}.}
 This analogue signal was then digitised using a multi-channel analyser from FastComTec ~\cite{fastcomtec}.
 The high voltage was provided using a  {CAEN~\cite{CAEN}} high voltage power supply {, model NDT1470}. The  {resulting LaBr$_{3}$ monitor} was calibrated using different gamma sources.

Measurements were performed using an AmBe neutron source, to verify sensitivity to the 480 keV gamma emissions from Boron in a high $\gamma$ background environment. The source is placed in  {a} polyethylene moderator to moderate the emitted fast neutrons~\cite{sf}. Good sensitivity over background was observed with the detector.

 {Next, this measurement was performed at a neutron beamline using the same equipment. The time-stamped option on the multi-channel analyser electronics was enabled to be able to measure the time-of-flight of the signal.}
The V20 beamline˜\cite{WORACEK2016} at the BERII research reactor at Helmholtz-Zentrum Berlin(HZB) provides a complete wavelength frame multiplication chopper system and is designed to replicate the ESS pulse
time structure~\cite{tblms}.



The source chopper was set to mimic the ESS pulse (2.86 ms pulse length with a repetition rate of 14 Hz) and the wavelength band choppers were set to prevent frame overlap within the repetition rate.
All other choppers were left open.
 { The opening of the source chopper was used to set the T0 of the time-of-flight signal.}
The integrated neutron flux is determined˜ {\cite{Maulerova2020} as $3\times10^6 \ n.cm^{-2}.s^{-1}$ in this configuration}.
The beam size can be collimated in both horizontal and vertical direction using several sets of slits in the beam line.
 {The beam size was set so that it fitted within the size of the chopper neutron window.}

\begin{figure}
 \centering
\includegraphics[width=80mm]{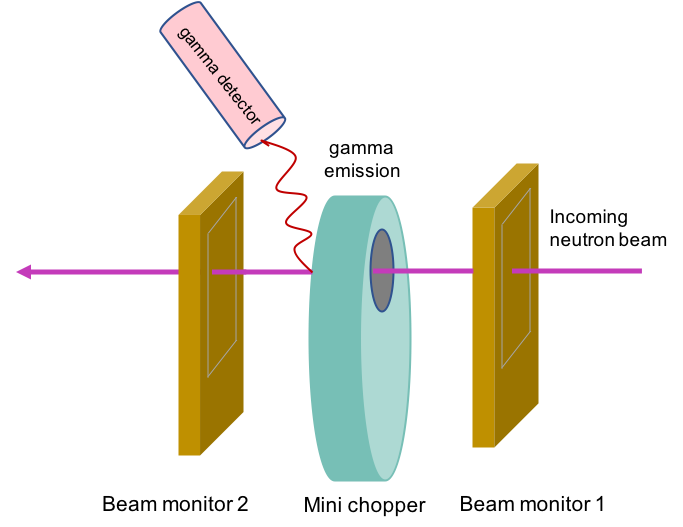}
\caption{Schematic of the experimental setup used on the V20 beamline in Berlin.
\label{Cartoon}}
\end{figure}

\begin{figure}
 \centering
\includegraphics[width=50mm]{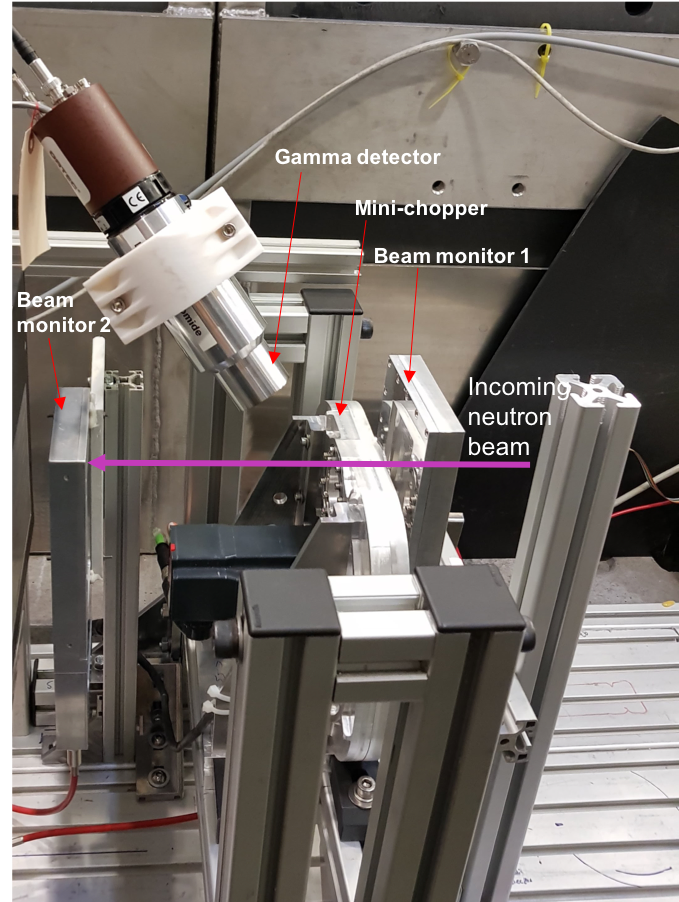}
\caption{Photograph of the experimental setup on the V20 beamline in Berlin.
The mini chopper is in the beam.
One monitor is  placed  before and another monitor is after the chopper.
The  {LaBr$_{3}$ monitor}  is placed close to the mini chopper.
The neutron beam passes from right to left in the photograph.
\label{Chopper_setup}}
\end{figure}

For the purpose of this measurement  {the LaBr$_{3}$ gamma monitor (without any shielding)} was placed close to a mini chopper˜\cite{Maulerova2020} which  {itself} was placed  in the direction of the neutron beam. The outer diameter of the chopper disc is 175 mm and made of  3 cm thick aluminum and coated with Boron Carbide on both sides with a total thickness of 3.5 mm  in order to stop the unwanted neutrons. The chopper rotates at 14 Hz and it has two openings to allow the neutrons with a specific energy range to go through.
 {The two openings are 2 and 4 ms long at 14 Hz.}
Thus the neutrons have to pass first the  aluminum window which is  25 by 25 mm wide and 0.5 mm thick before they either pass through the disc opening or get absorbed by the Boron coating.
 {Two beam monitors are placed upstream and downstream of the chopper.
These beam monitors, from Mirrotron˜\cite{Mirrortron} and Ordela˜\cite{Ordela} respectively, both utilise Helium-3 gas as the neutron sensitive medium, are not position sensitive and have similar performance˜\cite{BMpaper}, with an efficiency of about 0.1\% for 2.4 \AA \ neutrons. }
A schematic of the experimental setup is shown in fig.~\ref{Cartoon}.
A photograph of the setup is shown in fig.~\ref{Chopper_setup}.

\FloatBarrier

\section{Experimental Results}

\begin{figure}
\includegraphics[width=88mm]{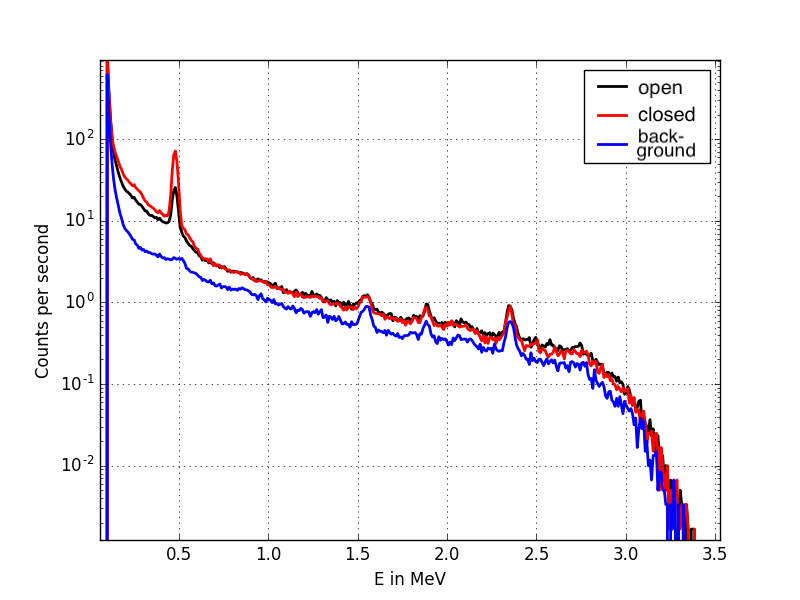}
\caption{Pulse height spectra for the  {LaBr$_{3}$ monitor} for background data (labelled background) and when the chopper disc is parked closed and parked open (labelled closed and open respectively).
\label{PHS_V20}}
\end{figure}

The pulse height spectra  were measured with the beam shutter closed and when the beam is on both with the chopper parked at the open position and at the closed position in fig.~\ref{PHS_V20}.
  {The 480 keV peak is clearly visible in these spectra.}
A deficit in the gamma counts is clearly visible around 480 keV when the chopper is parked open compared to the data when the chopper is parked closed {, despite the lack of shielding of the LaBr$_{3}$ monitor}.

\begin{figure}
\includegraphics[width=88mm]{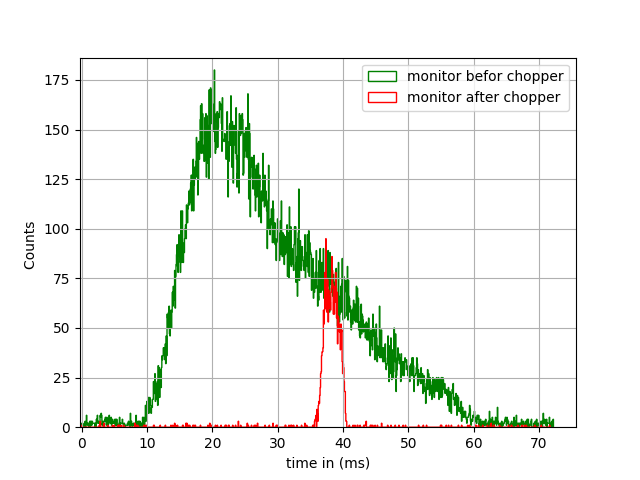}
\caption{Time-of-flight spectra of the monitor before and the monitor after the chopper disc.
\label{Mir_ORDELA}}
\end{figure}

The time of flight spectra of the monitor before and after the chopper measured with conventional beam monitors are shown in fig.~\ref{Mir_ORDELA}.
This shows the opening of the chopper is between 35 and 40 ms during the chopper rotation at 14 Hz.

\begin{figure}
 \includegraphics[width=88mm]{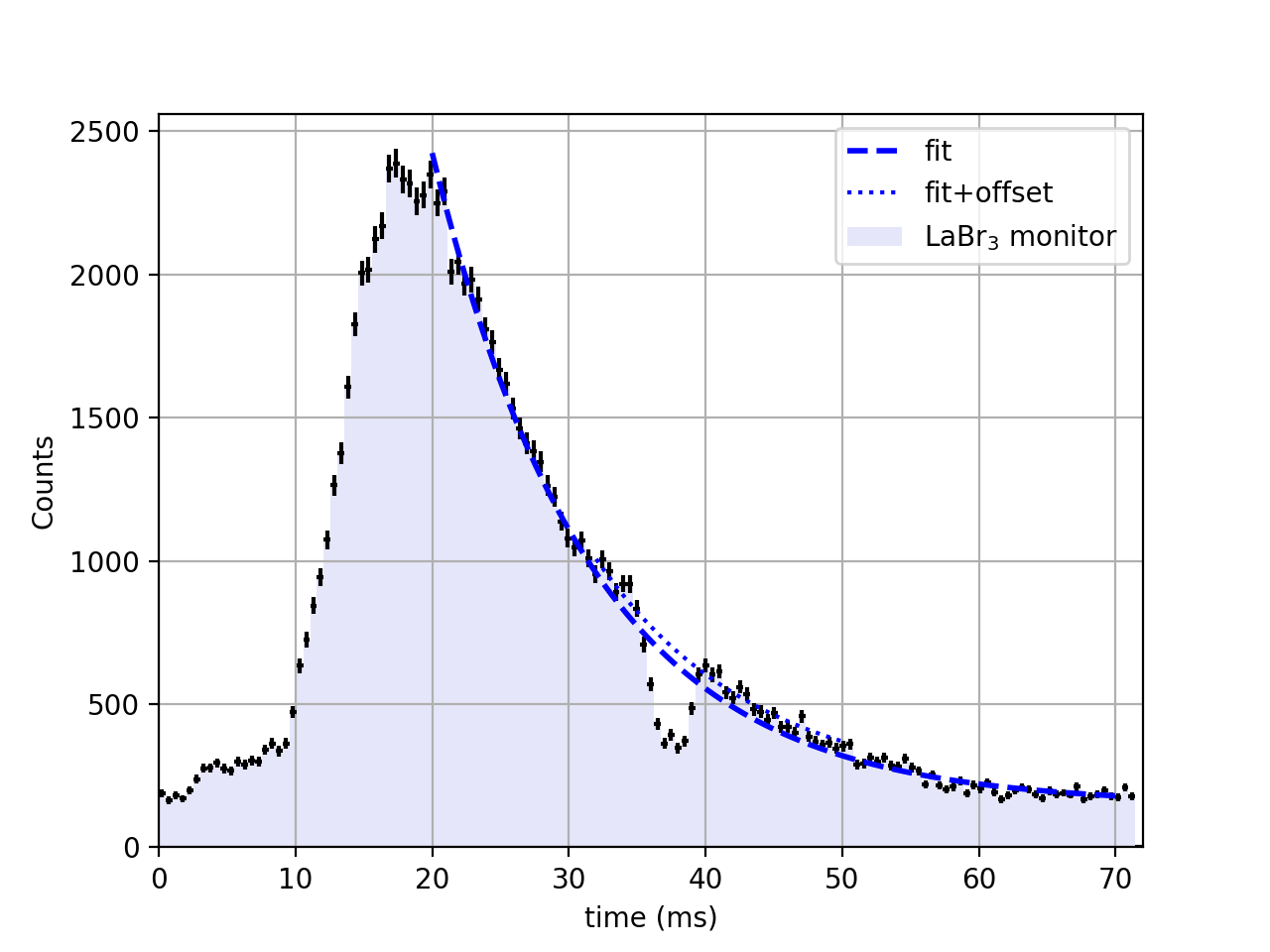}
\caption{Time-of-flight spectra from the  {LaBr$_{3}$} detector showing the deficit in events detected when the chopper slit was open.  The chopper disc was running at 14 Hz.
 {The dashed line is a fit (details in text) between 20 and 70 ms.
The second curve shown (dotted line) is this same fit offset by 50 counts to describe the trend of the data around the region of interest.}
\label{gamma_time}}
\end{figure}

\begin{figure}
   \includegraphics[width=88mm]{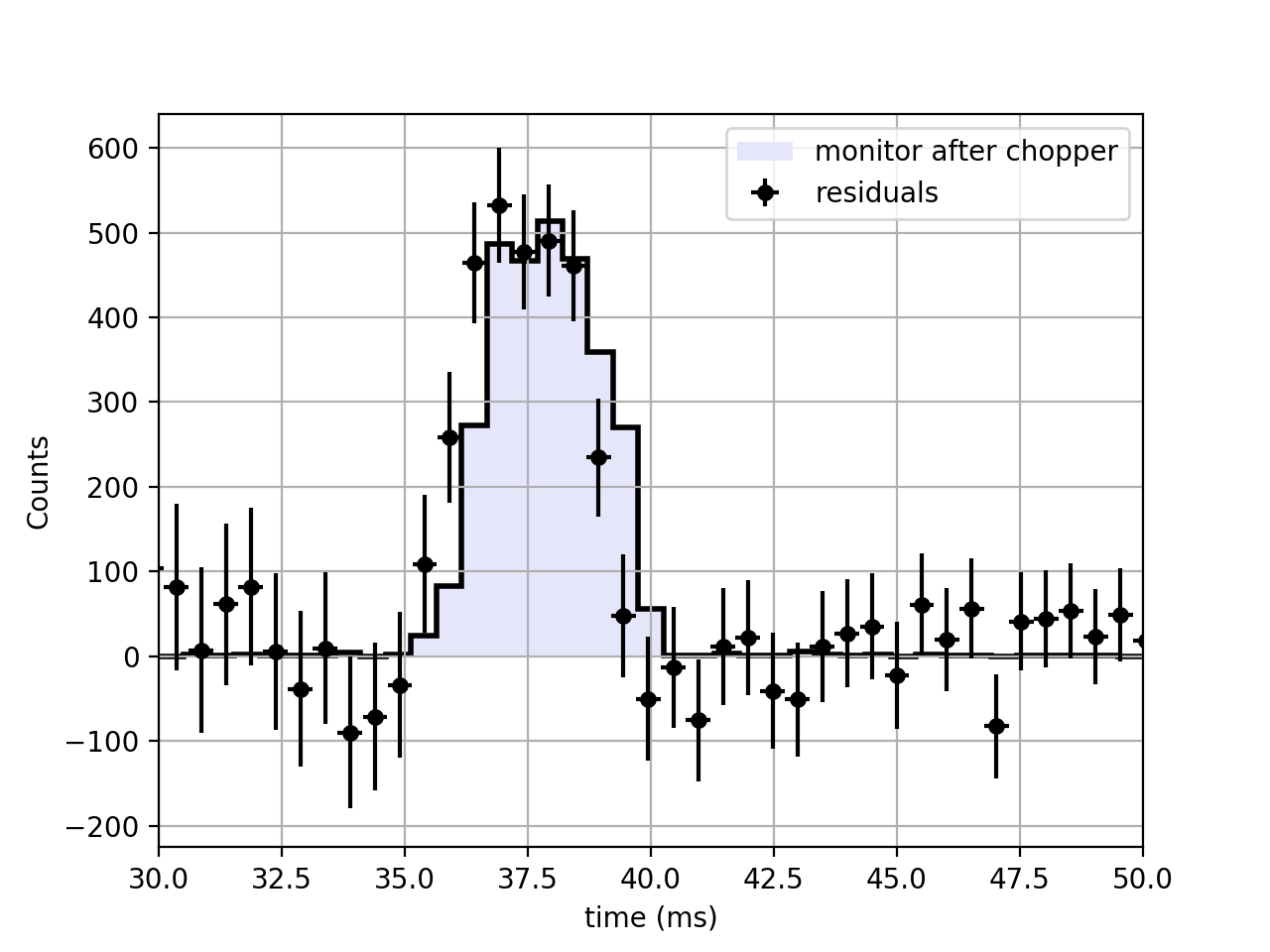}
\caption{
 {The time-of-flight spectra between 30 and 50 ms.
The solid line shows the data from the monitor after the chopper.
The points are the residuals between the fit and the LaBr$_{3}$ data, as shown in fig.˜\ref{gamma_time}.
The residuals have been  normalised to the peak of the counts from the monitor after the chopper.}
\label{fit}}
\end{figure}

The gammas detected in the  {LaBr$_{3}$} detector with energy within 480 $\pm$ 20 keV  were selected and plotted in time-of-flight as shown in fig.~\ref{gamma_time}.
This is measured with the chopper rotating at 14 Hz.
 {The data observed has a good correspondance with the pulse shape as shown by the monitor upstream of the chopper, as can be seen in fig.˜\ref{Mir_ORDELA}.
The time-of-flight spectra from the  {LaBr$_{3}$} detector show a clear decrease in counts between 35 ms and 40 ms and this time corresponds to the opening of the chopper as indicated by the monitor after the chopper, as shown in fig.˜\ref{Mir_ORDELA}.
This decrease in counts is expected and indicates the neutrons which pass through the chopper.
Where signal is observed in the LaBr$_{3}$ monitor, it indicates that this timeslice of the beam is absorbed in the chopper.

The data in fig.~\ref{gamma_time} has been fitted between 20 and 70 ms with an exponential plus linear function.
This fit provides a generally good description of this data, though it should be noted that the choice of function is somewhat arbitrary.
The data just around the signal of interest - between 32-35 ms and between 40-45 ms are slightly above the fit.
As shown by the second fit in fig.~\ref{gamma_time}, an offset of 50 counts to the previous fit provides a good description of the periphery to the region of interest.
This fit, including this additional offset, is used to construct the residuals, which is defined as the result of subtracting the number of the counts in the LaBr$_{3}$ data from this fit value.

Fig.~\ref{fit} shows a close-up of the results of these residuals. They are shown alongside the time-of-flight spectra from the monitor downstream of the chopper.
In Fig.~\ref{fit}  the residuals have been peak-normalised to the counts from the beam monitor downstream of the chopper.
The start of the pulse is ca. 0.5 ms later in the conventional beam monitor downstream of the chopper compared to the LaBr$_{3}$ monitor.
Additionally, the signal from beam monitor downstream of the chopper is also approximately 0.5 ms longer (i.e. the falling edge is 1 ms after the falling edge of these residuals). 
The flat ``top'' of both the signals is roughly equal in duration, and it is the shallower gradient of the falling edge which causes this effect.

The difference in the start of the pulse is explained because, whilst in the LaBr$_{3}$ scintillator the gamma which is emitted from the Boron in the chopper, is detected promptly, however in contrast, the neutron beam monitor is downstream of the chopper, as seen in Fig 4, by about 30-40 cm, and therefore is delayed.
The neutron wavelength of the neutrons selected by the chopper is roughly 5\AA, which implies a distance of about 37 cm for 500$\mu$s. 
Similar effects can been seen, for example in˜\cite{Dian2018, Backis2020}.

The difference in the falling edge for the downstream neutron beam monitor has two reasons.
Firstly, there is a wavelength band allowed through the chopper. Slower neutrons will take longer to travel to the chopper, and this wavelength spread will spread out the signal correspondingly.
Additionally there will be effects from scattering from upstream beamline components, and also reflected neutrons from components after this chopper, which will affect the timing of the measurement.
}

This feature proves the feasibility of this method to be used for phasing of the chopper without the addition of any material in the beamline.

\section{Conclusions}
A  {LaBr$_{3}$} gamma detector was placed close to a mini chopper on a pulsed beamline with the chopper running.
A clear time-variant gamma peak  at  480 keV was detected.
This gamma emission is due to the interaction of the incoming neutrons with  Boron coating on the chopper system .
A remarkable dip in the gamma at 480 keV curve was observed indicating the time of opening and closing of the chopper.
This is a proof of concept of the parasitic method for monitoring the beam nearby a chopper system.
The results applied here are equally applicable to choppers using Gadolinium absorbers, though a different, higher, energy window would need to be used.
This concept, developed into an engineered system, could be a good diagnostic tool without adding any material in the beam, i.e. a parasitic beam monitor.
It is particularly appropriate for determining and verifying chopper phases {, as the measurement is prompt with the interaction of the neutron with the chopper disk, with no timing uncertainties introduced from corrections due to monitor location, wavelength band of the neutrons or scattering from beamline materials}.
 {A real implementation on a beamline should encompass an effective shielding design to select gammas originating from the chopper and reject those from other beamline sources to enhance the signal to background of the measurement.}

\acknowledgments
This work was funded by EU Horizon2020 framework, BrightnESS project 676548.
Experimental measurements were carried out at the Source Testing Facility, Lund University (LU - Sweden) and at the V20 beamline at HZB, Berlin (DE).
The authors would like to acknowledge R.Woracek for support during the beam line test at HZB.
\medskip
\bibliographystyle{eplbib.bst}

\end{document}